\documentclass[aps,pra,reprint,amsfonts,amssymb,superscriptaddress,floatfix,eprint]{revtex4-1}
\usepackage{graphicx}
\usepackage{amsmath}
\usepackage{dcolumn}
\usepackage{color}
\usepackage{float}
\usepackage{calc}
\usepackage{MnSymbol}
\usepackage{hyperref}
\usepackage{hypernat}

\newcommand{\edd}{\epsilon_\text{dd}}

\begin{document}

\title{Vortex Lattice Formation in Dipolar Bose-Einstein Condensates via Rotation of the Polarization}

\author{Srivatsa B. Prasad}
\email{srivatsa.badariprasad@unimelb.edu.au}
\affiliation{School of Physics, University of Melbourne, Melbourne, 3010, Australia}
\author{Thomas Bland}
\affiliation{Joint Quantum Centre Durham-Newcastle, School of Mathematics, Statistics and Physics, Newcastle University, Newcastle upon Tyne, NE1 7RU, United Kingdom}
\author{Brendan C. Mulkerin}
\affiliation{Centre for Quantum and Optical Science, Swinburne University of Technology, Melbourne, 3122, Australia}
\author{Nick G. Parker}
\affiliation{School of Physics, University of Melbourne, Melbourne, 3010, Australia}
\affiliation{Joint Quantum Centre Durham-Newcastle, School of Mathematics, Statistics and Physics, Newcastle University, Newcastle upon Tyne, NE1 7RU, United Kingdom}
\author{Andrew M. Martin}
\affiliation{School of Physics, University of Melbourne, Melbourne, 3010, Australia}

\date{\today}

\begin{abstract}
The behaviour of a harmonically trapped dipolar Bose-Einstein condensate with its dipole moments rotating at angular frequencies lower than the transverse harmonic trapping frequency is explored in the co-rotating frame. We obtain semi-analytical solutions for the stationary states in the Thomas-Fermi limit of the corresponding dipolar Gross-Pitaevskii equation and utilise linear stability analysis to elucidate a phase diagram for the dynamical stability of these stationary solutions with respect to collective modes. These results are verified via direct numerical simulations of the dipolar Gross-Pitaevskii equation, which demonstrate that dynamical instabilities of the co-rotating stationary solutions lead to the seeding of vortices that eventually relax into a triangular lattice configuration. Our results illustrate that rotation of the dipole polarization represents a new route to vortex formation in dipolar Bose-Einstein condensates.
\end{abstract}

\maketitle
\section{\label{sec:level0}Introduction}
Interacting quantum gases have proved to be a fertile testing ground for theories of many-body physics in recent years. In particular, dipolar quantum gases such as dipolar Bose-Einstein condensates (BECs) and degenerate dipolar fermionic gases offer a unique route to novel many-body phenomena where the subtle interplay between long- and short-range interactions plays a significant role~\cite{physrep_464_3_71-111_2008, repprogphys_72_12_126401_2009, chemrev_112_9_5012-5061_2012}, and considerable progress has been made in the study of dipolar BECs in the last two decades. After the first dipolar BEC was produced using chromium in 2005~\cite{prl_94_16_160401_2005}, the effects of the dipolar interaction on the stability and excitations of the condensate were of much interest, with particular focus on the anisotropy of the condensate's superfluidity~\cite{prl_101_8_080401_2008, natphys_4_3_218-222_2008, prl_121_3_030401_2018} and the theoretical possibility of roton modes and supersolidity~\cite{prl_90_25_250403_2003, prl_98_3_030406_2007}. More recently the production of dysprosium and erbium BECs, which feature considerably stronger dipolar interactions than Cr BECs, has led to the discovery of an increasingly diverse range of exotic phenomena~\cite{prl_107_19_190401_2011, prl_108_21_210401_2012}. Examples of these include self-bound dipolar droplets~\cite{nature_530_7589_194-197_2016, nature_539_7628_259-262_2016, prl_116_21_215301_2016, prx_6_4_041039_2016} and the experimental confirmation of the presence of roton modes~\cite{natphys_14_5_442-446_2018, prl_122_18_183401_2019} and the supersolid phase~\cite{prl_122_13_130405_2019, prx_9_1_011051_2019, prx_9_2_021012_2019}. Many of these have been found to be explicitly dependent on the role of beyond-mean-field quantum fluctuations in stabilizing a strongly dipolar BEC~\cite{pra_86_6_063609_2012, pra_94_3_033619_2016}.

One area of considerable research in BECs has centred on the emergence and properties of vortices, whose circulations are necessarily quantised due to the condensate wavefunction being locally single-valued~\cite{science_292_5516_476-479_2001, prl_91_10_100402_2003, prl_103_4_045301_2009, prl_117_24_245301_2016, science_364_6447_1264-1267_2019, science_364_6447_1267-1271_2019}. Examples of the methods that experimentalists have used to create vortices include: direct imprinting of phase defects into the condensate~\cite{prl_89_19_190403_2002}, rotation of either a laser beam stirrer or the external trapping potential of the condensate itself~\cite{prl_84_5_806-809_2000, prl_88_1_010405_2001}, dragging a barrier through the condensate~\cite{prl_104_16_160401_2010, science_364_6447_1264-1267_2019, science_364_6447_1267-1271_2019}, applying a rapidly oscillating perturbation to the trapping potential~\cite{prl_103_4_045301_2009}, Bose-condensing a rotating normal Bose gas~\cite{prl_87_23_210403_2001}, and utilising the Kibble-Zurek mechanism to trigger the formation of topological defects~\cite{nature_455_7215_948-951_2008}. While vortices have not yet been experimentally observed in dipolar BECs, there exists an extensive body of theoretical research regarding vortex structure, vortex lattice structure, and vortex-vortex interactions in these systems~\cite{prl_95_20_200402_2005, prl_95_20_200403_2005, pra_73_6_061602r_2006, pra_75_2_023623_2007, prl_100_24_240403_2008, pra_79_21_063622_2009, prl_111_17_170402_2013, jphysb_49_15_155301_2016, pra_98_2_023610_2018, pra_98_2_023618_2018, prl_121_17_174501_2018, jphyscondesmatter_29_10_103004_2017}.

In this work we propose a new mechanism for producing vortices in dipolar BECs that involves the direct rotation of the polarizing field of the dipoles. For a dipolar BEC in a harmonic trap with an initially static dipole polarization we consider the effects of slowly increasing the rotation frequency of the dipole polarization about an orthogonal axis. Using a semi-analytical model based on dipolar Gross-Pitaevskii theory, it is shown that the co-rotating stationary state of the condensate becomes dynamically unstable against symmetry-breaking collective modes as the rotation frequency approaches the transverse trapping frequency. Numerically, it is demonstrated that this dynamical instability results in the formation of vortices in the previously vorticity-free condensate, and that these vortices ultimately self-order into a triangular Abrikosov vortex lattice. Experimentally, the rotation of the polarization of a dipolar BEC has already been achieved in the context of tuning the effective strength of the time-averaged dipole-dipole interaction~\cite{prl_89_13_130401_2002, prl_120_23_230401_2018}. Crucially, the rotation frequency in such studies is at least two orders of magnitude larger than that investigated in our work, suggesting that the mechanism we propose for the production of vortices in dipolar BECs is readily achievable via the use of existing technologies.

This paper is structured as follows. In Section~\ref{sec:level1} we discuss a semi-analytical formalism, based on the dipolar Gross-Pitaevskii equation and its reformulation as the equations of dipolar superfluid hydrodynamics, which we use to solve self-consistently for the Thomas-Fermi stationary solutions of a dipolar BEC with its dipoles rotating perpendicularly to their alignment. These solutions are presented in detail in Section~\ref{sec:level2}. In Section~\ref{sec:level3}, we use the dipolar Gross-Pitaevskii equation to numerically simulate a dipolar BEC with the dipole rotation frequency being slowly increased from zero. and demonstrate that while the condensate initially obeys the Thomas-Fermi stationary solution, it develops an instability that results in vortex formation as the rotation frequency approaches the transverse trapping frequency. Finally in Section~\ref{sec:level4}, it is shown via linearization of the time-dependent superfluid hydrodynamic equations that the Thomas-Fermi stationary solutions are dynamically unstable against collective modes in the regime in which vortex formation is observed in Section~\ref{sec:level3}. 

\section{\label{sec:level1}Thomas-Fermi Theory for the Rotating Dipolar BEC}
Our semi-analytical description of the dipolar BEC with rotating dipole moments begins with a summary of the mean-field theory that we shall employ throughout this paper. We consider a dilute condensate whose bosons have intrinsic magnetic or electric dipoles that are aligned along an externally applied polarizing field. We utilise the notion of the mean-field order parameter, $\psi \equiv \langle\hat{\psi}_0\rangle$, which is defined as the vacuum expectation value of the bosonic annihilation operator and is normalized such that $\int\mathrm{d}\mathbf{r}|\psi(\mathbf{r},t)|^2$ equals $N$, the number of condensed bosons. At the lowest order of approximation, the dipolar BEC is described by the dipolar Gross-Pitaevskii theory~\cite{physrep_464_3_71-111_2008, repprogphys_72_12_126401_2009, chemrev_112_9_5012-5061_2012} -- a classical field theory for $\psi$. In a reference frame that rotates at an angular frequency $\mathbf{\Omega} = \Omega\hat{z}$ with respect to the laboratory frame, the dipolar Gross-Pitaevskii equation (dGPE) is given by
\begin{equation}
i\hbar\frac{\partial\psi}{\partial t} = \left[-\frac{\hbar^2\nabla^2}{2m} + V_{\text{T}} + V_{\text{int}} + i\hbar\Omega\left(x\frac{\partial}{\partial y}-y\frac{\partial}{\partial x}\right)\right]\psi.  \label{eq:dgpe}
\end{equation}
where $m$ is the mass of each dipole. Here we consider the external trapping potential $V_{\text{T}}$ to be a harmonic trap of the form
\begin{equation}
V_{\text{T}}\left(\mathbf{r}\right) = \frac{1}{2}m\left[\omega_{\perp}^2\rho^2 + \omega_z^2z^2\right], \label{eq:trap}
\end{equation}
The axial symmetry of the trapping about $\hat{z}$ is imposed so that the trapping is identical in both the rotating and laboratory frames, and we also define the axial trapping ratio $\gamma = \omega_z/\omega_{\perp}$. The relevant interactions between the dipoles are the short-range, isotropic, van der Waals interaction and the anisotropic, long-range, dipole-dipole interaction. These are accounted for in the dGPE by a two-body interaction term, $V_{\text{int}}$~\cite{pra_61_5_051601r_2000, prl_85_9_1791-1794_2000, pra_63_5_053607_2001}:
\begin{equation}
V_{\text{int}}(\mathbf{r},t) = g|\psi|^2 + \frac{C_{\text{dd}}}{4\pi}\int\mathrm{d}\mathbf{r'}\,\frac{1-3\cos^2\theta_{\mathbf{r},\mathbf{r}'}(t)}{|\mathbf{r}-\mathbf{r}'|^3}\left|\psi(\mathbf{r'},t)\right|^2. \label{eq:ddpot}
\end{equation}
Here the interaction strength of the local contribution is given by $g = 4\pi\hbar^2a_{\text{s}}/m$, where $a_{\text{s}}$ is the $s$-wave scattering length of the relevant two-body scattering potential, while the angle $\theta_{\mathbf{r},\mathbf{r}'}$ is defined through the relation $\cos^2\theta_{\mathbf{r},\mathbf{r}'}(t) = \left[\hat{e}(t)\cdot(\hat{\mathbf{r}}-\hat{\mathbf{r}}')\right]^2$, where $\hat{\mathbf{e}}(t)$ is the polarization direction of the dipoles. We also define the interaction ratio $\edd = C_{\text{dd}}/(3g)$ and for the later ease of use, we also state an equivalent definition of $V_{\text{int}}$ in terms of a dipolar pseudopotential $\phi_{\text{dd}}$,
\begin{gather}
V_{\text{int}}\left(\mathbf{r},t\right) = g(1-\edd)n\left(\mathbf{r},t\right)-3g\edd\left(\hat{\mathbf{e}}\cdot\nabla\right)^2\phi_{\text{dd}}\left(\mathbf{r},t\right), \label{eq:ddpotfinal} \\
\phi_{\text{dd}}\left(\mathbf{r},t\right) = \frac{1}{4\pi}\int\mathrm{d}^3r'\,\frac{\left|\psi\left(\mathbf{r}',t\right)\right|^2}{\left|\mathbf{r}-\mathbf{r}'\right|}, \label{eq:pseudopot}
\end{gather}
such that $\nabla^2\phi_{\text{dd}} = -4\pi|\psi|^2$~\cite{prl_92_25_250401_2004, pra_71_3_033618_2005}. In this work, we assume that $\hat{\mathbf{e}}$ rotates in the $x$-$y$ plane about $\hat{z}$ at the angular frequency $\Omega$ and, as such, we fix $\hat{\mathbf{e}} = \hat{x}\,\forall\,t$ in the co-rotating frame without loss of generality.

The form of the interaction strength $C_{\text{dd}}$ is dependent on whether the polarizing field is electric or magnetic. For a species with electric dipole moment $d$, $C_{\text{dd}} = d^2/\epsilon_0$, where $\epsilon_0$ is the permittivity of free space, while the expression for a species with magnetic dipole moment $\mu$ is $C_{\text{dd}} = \mu_0\mu^2$, where $\mu_0$ is the permeability of free space. In terms of commonly used dipolar BEC species, the bare value of $\edd = C_{\text{dd}}/(3g)$ ranges from $0.16$ for $^{52}$Cr~\cite{prl_97_25_250402_2006} to as high as $1.42$ for $^{164}$Dy~\cite{pra_92_2_022703_2015}, both of which are orders of magnitude higher than those of `nondipolar' species such as $^{87}$Rb~\cite{repprogphys_72_12_126401_2009}. However, $\edd$ represents an additional tuneable parameter alongside $\Omega$ and $\gamma$ in this system due to the ability to tune $g$ experimentally by exploiting Feshbach resonances~\cite{prl_101_8_080401_2008, natphys_4_3_218-222_2008}. Recent theoretical studies of dipolar BECs, motivated by experimental findings highlighting the role of beyond-mean-field effects in the behaviour of strongly dipolar BECs ($\edd \gtrsim 1$) have included an additional term proportional to $|\psi|^5$ on the right hand side of Eq.~\eqref{eq:dgpe}~\cite{pra_86_6_063609_2012, pra_94_3_033619_2016, prl_121_19_195301_2018}. However, our analysis is restricted to values of $\edd$ considerably less than $1$, and as such, we expect that Eq.~\eqref{eq:dgpe} accurately describes the behaviour of such a system in the dilute limit.

To solve Eq.~\eqref{eq:dgpe} we shall re-express the order parameter $\psi$ as $\psi = \sqrt{n}\exp{(iS)}$, where $n = |\psi|^2$ and $S$ are interpreted as the condensate's number density and phase respectively. By applying this transformation to Eq.~\eqref{eq:dgpe} and separating out the resulting real and imaginary terms, we obtain the dipolar superfluid hydrodynamic equations~\cite{prl_86_3_377-390_2001, prl_87_19_190402_2001, prl_92_25_250401_2004}:
\begin{align}
\frac{\partial n}{\partial t} &= -\nabla\cdot\left[n\left(\mathbf{v}-\mathbf{\Omega}\times\mathbf{r}\right)\right], \label{eq:continuity} \\
m\frac{\partial\mathbf{v}}{\partial t} &= -\nabla\left[\frac{1}{2}m\mathbf{v}^2-m\mathbf{v}\cdot\left(\mathbf{\Omega}\times\mathbf{r}\right)+V_{\text{T}}+V_{\text{int}}\right] \nonumber \\
&+ \nabla\left[\frac{\hbar^2}{2m\sqrt{n}}\nabla^2(\sqrt{n})\right]. \label{eq:euler}
\end{align}
Here the condensate's velocity field, $\mathbf{v}$,is defined as
\begin{equation}
\mathbf{v} = \frac{\hbar\nabla S}{m}. \label{eq:velocity}
\end{equation}
Equation~\eqref{eq:continuity} is analogous to the continuity equation for a classical fluid, while Eq.~\eqref{eq:euler} is analogous to the Euler equation for inviscid fluid flow. In the Thomas-Fermi (TF) regime, which generally exists when $Na_s$ is sufficiently high, the effects of zero-point kinetic energy fluctuations of the condensate are negligible~\cite{pra_51_2_1382-1386_1995, pra_78_4_041601r_2008}. This is equivalent to ignoring the `quantum pressure' term in Eq.~\eqref{eq:euler} that is proportional to $\nabla[\nabla^2(\sqrt{n})/\sqrt{n}]$, with the resulting simplified Euler equation given by
\begin{equation}
m\frac{\partial\mathbf{v}}{\partial t} = -\nabla\left[\frac{1}{2}m\mathbf{v}^2-m\mathbf{v}\cdot\left(\mathbf{\Omega}\times\mathbf{r}\right)+V_{\text{T}}+V_{\text{int}}\right]. \label{eq:tfeuler}
\end{equation}
When $\Omega = 0$, the rotating frame is equivalent to the laboratory frame, with Eqs.~\eqref{eq:continuity} and \eqref{eq:tfeuler} reducing to the description of a dipolar BEC in the TF regime with a time-independent polarization axis, a limit that has been documented extensively in previous studies~\cite{prl_92_25_250401_2004, pra_71_3_033618_2005}.

Initially, we focus on the stationary solutions of Eqs.~\eqref{eq:dgpe} and~\eqref{eq:tfeuler}, whose time dependence is specified by the condensate's chemical potential, $\mu$, as
\begin{equation}
\psi(\mathbf{r},t) = \psi(\mathbf{r}, t = 0)\exp\left(\frac{-i\mu t}{\hbar}\right). \label{eq:chempot}
\end{equation}
Via the definitions of $n$ and $v$, and Eqs.~\eqref{eq:ddpotfinal},~\eqref{eq:pseudopot},~\eqref{eq:continuity},~\eqref{eq:tfeuler} and~\eqref{eq:chempot}, the stationary solutions in the TF limit obey
\begin{align}
0 &= \nabla\cdot\left[n\left(\mathbf{v} - \mathbf{\Omega}\times\mathbf{r}\right)\right], \label{eq:continuitystat} \\
\mu &= \frac{1}{2}m\mathbf{v}\cdot\left[\mathbf{v} - 2\left(\mathbf{\Omega}\times\mathbf{r}\right)\right] + V_{\text{T}} \nonumber \\
&+ g(1 - \edd)n - 3g\edd\frac{\partial^2\phi_{\text{dd}}}{\partial x^2}. \label{eq:tfeulerstat}
\end{align}
Let us impose \textit{Ans{\"a}tze} for $n$ and $S$ of the form
\begin{align}
n_{\text{TF}}(\mathbf{r}) &= n_0\left(1-\frac{x^2}{R_x^2}-\frac{y^2}{R_x^2}-\frac{z^2}{R_z^2}\right), \label{eq:tfdensity} \\
S_{\text{TF}}(\mathbf{r}, t) &= (m\alpha xy - \mu t)/\hbar, \label{eq:tfphase}
\end{align}
and define the ratios $\kappa_x = R_x/R_z$ and $\kappa_y = R_y/R_z$. Here the TF density is defined as nonzero only when the quantity expressed in Eq.~\eqref{eq:tfdensity} is positive, while $n_0 = 15N/(8\pi R_xR_yR_z)$ is a factor that normalises $n_{\text{TF}}$ to $N$. Exact solutions within the TF approximation may be determined via a set of self-consistency relations that are satisfied by the parameters $\kappa_x$, $\kappa_y$, $R_z$ and $\alpha$, which we shall proceed to outline briefly. Firstly, via Eq.~\eqref{eq:velocity}, the choice of $S_{\text{TF}}$ implies that the resulting velocity field, $\mathbf{v} = \alpha\nabla(xy)$, has zero vorticity. Substituting Eqs.~\eqref{eq:tfdensity} and~\eqref{eq:tfphase} into Eq.~\eqref{eq:continuitystat} results in the condition~\cite{prl_86_3_377-390_2001}
\begin{equation}
\alpha = \left(\frac{\kappa_x^2-\kappa_y^2}{\kappa_x^2+\kappa_y^2}\right)\Omega. \label{eq:alphaomegadefn}
\end{equation}
Equation~\eqref{eq:alphaomegadefn} provides an intuitive meaning for the amplitude of the velocity field, $\alpha$: for positive nonzero $\Omega$ the cross-section of the condensate's density profile, in the $x$-$y$ plane, is elongated along the co-rotating $x$-axis ($y$-axis) when $\alpha$ is positive (negative). Thus the condensate is axially symmetric about $\hat{z}$ when $\alpha = 0$.

The dipolar pseudopotential, $\phi_{\text{dd}}$, that corresponds to the TF density specified in Eq.~\eqref{eq:tfdensity} is exactly given by~\cite{pra_71_3_033618_2005, pra_82_3_033612_2010, prl_122_5_050401_2019}
\begin{align}
\phi_{\text{dd}}(x,y,z) &= \frac{n_0\kappa_x\kappa_y}{4}\left(\frac{\beta_{000}}{2} - x^2\beta_{100}-y^2\beta_{010}-z^2\beta_{001}\right) \nonumber \\
&+ \frac{n_0\kappa_x\kappa_y}{8R_z^2}(x^4\beta_{200} + y^4\beta_{020}+z^4\beta_{002}) \nonumber \\
&+ \frac{n_0\kappa_x\kappa_y}{4R_z^2}(x^2y^2\beta_{110}+y^2z^2\beta_{011}+x^2z^2\beta_{101}), \label{eq:pseudopotbeta}
\end{align}
where $\beta_{ijk}\left(\kappa_x,\kappa_y\right)$ denotes the following integral:
\begin{equation}
\beta_{ijk}\left(\kappa_x,\kappa_y\right) = \int_0^{\infty}\frac{\mathrm{d}s}{(\kappa_x^2+s)^{i+\frac{1}{2}}(\kappa_y^2+s)^{j+\frac{1}{2}}(1+s)^{k+\frac{1}{2}}}. \label{eq:beta}
\end{equation}
If we define the effective $x$ and $y$ trapping frequencies as
\begin{align}
\widetilde{\omega}_x^2 &= \omega_{\perp}^2 + \alpha^2 - 2\alpha\Omega, \label{eq:omegax} \\
\widetilde{\omega}_y^2 &= \omega_{\perp}^2 + \alpha^2 + 2\alpha\Omega, \label{eq:omegay}
\end{align}
evaluating Eq. \eqref{eq:tfeulerstat} gives us the self-consistency relations we seek~\cite{prl_122_5_050401_2019}:
\begin{gather}
\kappa_x^2 = \frac{1}{\zeta}\left(\frac{\omega_{\perp}\gamma}{\widetilde{\omega}_x}\right)^2\left[1 + \edd\left(\frac{9}{2}\kappa_x^3\kappa_y\beta_{200}-1\right)\right], \label{eq:kappax} \\
\kappa_y^2 = \frac{1}{\zeta}\left(\frac{\omega_{\perp}\gamma}{\widetilde{\omega}_y}\right)^2\left[1 + \edd\left(\frac{3}{2}\kappa_y^3\kappa_x\beta_{110}-1\right)\right], \label{eq:kappay}  \\
R_z^2 = \frac{2gn_0}{m\gamma^2\omega_{\perp}^2}\zeta, \label{eq:rz2} \\
\zeta = 1 + \edd\left(\frac{3}{2}\kappa_x\kappa_y\beta_{101} - 1\right), \label{eq:zeta}
\end{gather}
\begin{align}
0 &= (\alpha + \Omega)\left[\widetilde{\omega}_x^2 - \frac{9}{2}\edd\frac{\omega_{\perp}^2\kappa_x\kappa_y\gamma^2}{\zeta}\beta_{200}\right] \nonumber \\
&+ (\alpha - \Omega)\left[\widetilde{\omega}_y^2 - \frac{3}{2}\edd\frac{\omega_{\perp}^2\kappa_x\kappa_y\gamma^2}{\zeta}\beta_{110}\right]. \label{eq:alphakappa}
\end{align}
To obtain the velocity field as well as the shape of the density it is sufficient to solve Eqs.~\eqref{eq:kappax},~\eqref{eq:kappay} and~\eqref{eq:alphakappa} self-consistently.

\section{\label{sec:level2}Stationary Solutions of the Thomas-Fermi Problem}
\begin{figure}
\includegraphics[width=\linewidth]{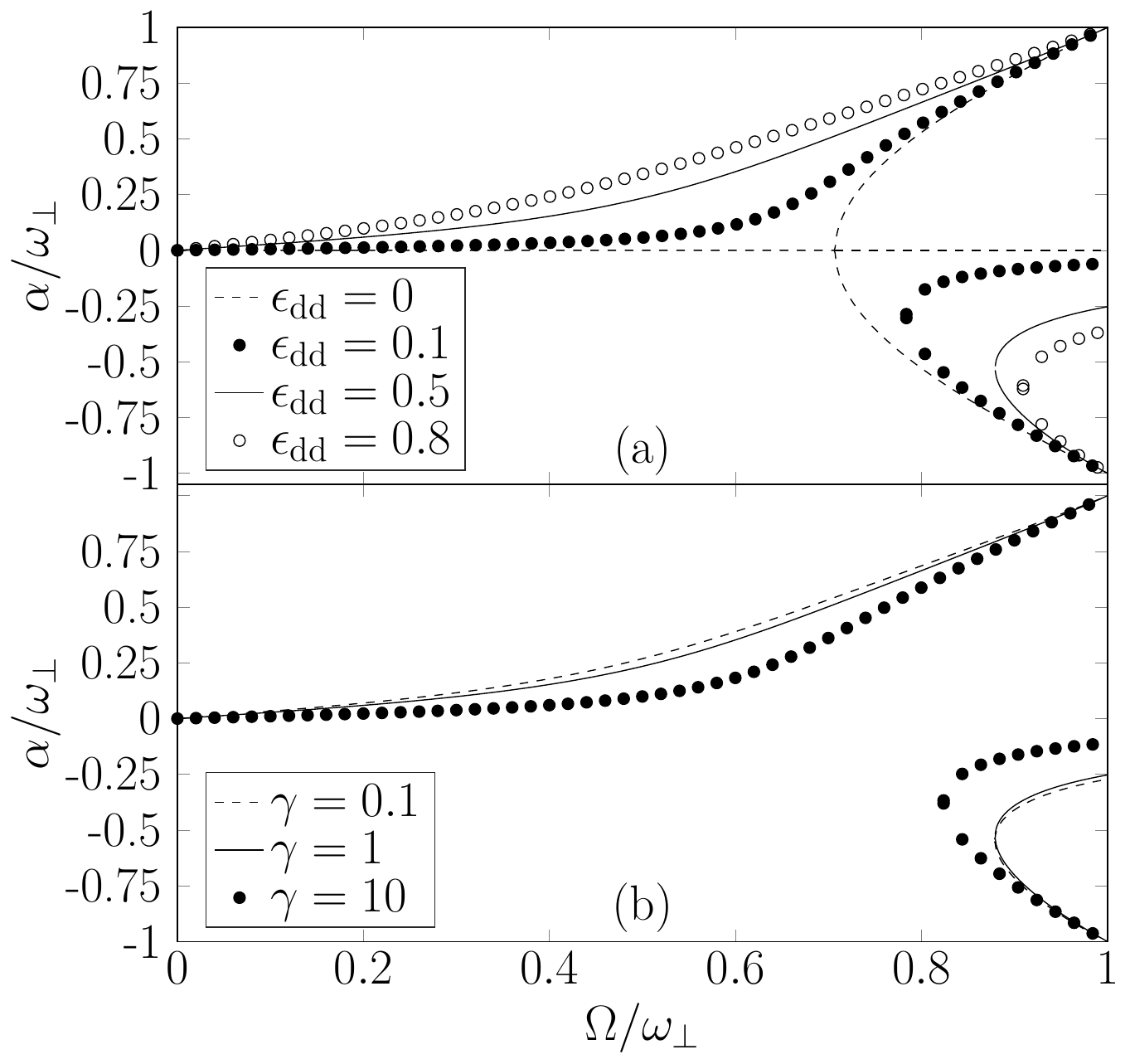}
\vspace*{-5mm}
\caption{Stationary solutions, as characterized by $\alpha$, as a function of $\Omega$: (a) $\gamma = 1$ and various $\edd$; (b) $\edd = 0.5$ and various $\gamma$. Branch I obeys $\alpha \geq 0$ while Branches II and III obey $\alpha \leq 0$, and $\Omega = \Omega_{\text{b}}$ when Branches II and III coincide.}
\label{fig:bothrangeomegaprofile}
\end{figure}

In Fig.~\ref{fig:bothrangeomegaprofile}, we plot $\alpha/\omega_{\perp}$ as a function of $\Omega/\omega_{\perp}$ at fixed $\gamma = 1$ for $\edd \in \lbrace 0, 0.1, 0.5, 0.8\rbrace$ (a) and at fixed $\edd = 0.5$ for $\gamma \in \lbrace 0.1, 1, 10\rbrace$ (b). In Ref.~\cite{prl_122_5_050401_2019}, it was found that a bifurcation exists in this system at a given rotation frequency $\Omega_{\text{b}}$, a quantity that is dependent on $\edd$ and $\gamma$, at which the number of stationary solutions increases from 1 to 3~\cite{prl_122_5_050401_2019}. When $\edd = 0$, the bifurcation diagram is symmetric about the $\Omega$ axis and two additional symmetric branches emerge when $\Omega = \omega_{\perp}/\sqrt{2}$. Let us denote the $\alpha > 0$ and $\alpha < 0$ branches as Branches I and III respectively; both of these \textit{subcritical} branches terminate at $\Omega = \omega_{\perp}$, a limit which is characterised by $\alpha\rightarrow \pm\omega_{\perp}$. However the $\alpha = 0$ branch, which we shall denote as Branch II, persists for all $\Omega$. Conversely, when $\edd > 0$, Branch I starts from $\alpha = 0$ at $\Omega = 0$ and exhibits a monotonic increase of $\alpha$ until the limit $\alpha\rightarrow\omega_{\perp}$ as $\Omega\rightarrow\omega_{\perp}$. Instead of the symmetric bifurcation that is present when $\edd = 0$, a nonzero $\edd$ results in a symmetry-breaking bifurcation where Branches II and III obey $\alpha < 0$ and are simply connected at $\Omega = \Omega_{\text{b}}(\edd, \gamma)$ but are disconnected from Branch I. While Branch III terminates at $\Omega = \omega_{\perp}$ with $\alpha\rightarrow -\omega_{\perp}$ as $\Omega\rightarrow\omega_{\perp}$, the \textit{overcritical} Branch II persists for $\Omega \rightarrow \infty$. Though we do not consider this limit, it has been found that Branch II monotonically approaches the limit $\alpha\rightarrow 0^{-}$ as $\Omega\rightarrow\infty$~\cite{prl_122_5_050401_2019}. In Fig.~\ref{fig:bothrangeomegaprofile} (b), it is seen that the effect of the trapping aspect ratio $\gamma$ is qualitative and leads to shifts in the bifurcation frequency relative to the case of the spherically symmetric trap. We also note that for Branch I, $\frac{\mathrm{d}\alpha}{\mathrm{d}\Omega}$ is higher for smaller $\gamma$ and higher $\epsilon_{\text{dd}}$.

The qualitative features of the bifurcation diagram shown in Fig.~\ref{fig:bothrangeomegaprofile} are similar to the diagram corresponding to the TF limit of a BEC, with or without $z$-polarised dipoles, subject to a harmonic trap that is anistropic in the $x$-$y$ plane and rotating about $\hat{z}$~\cite{prl_86_3_377-390_2001, pra_80_3_033617_2009}. However in such systems the symmetry-breaking parameter is not $\edd$ but is instead the $x$-$y$ trapping ellipticity. It is also possible to obtain stationary solutions and a corresponding phase diagram in the presence of the rotation of both an anisotropic trap and the dipole polarization. In this regime, the interplay between a positive $\edd$ with the dipoles polarized along $\hat{x}$ in the co-rotating frame and a negative trapping ellipticity -- that is, an elongation of the trap along the co-rotating $y$-axis -- may lead to interesting effects. However, as a minimal model of vortex nucleation in dipolar BECs we assume that the trapping is cylindrically symmetric about $\hat{z}$.

\section{\label{sec:level3}Dipolar Gross-Pitaevskii Equation Simulations}
In this system the properties of Branch II ($\alpha < 0\,\forall\,\Omega\in[\Omega_{\text{b}},\infty)$) have been studied in previous investigations into the effective rotational tuning of $\edd$~\cite{prl_122_5_050401_2019, arXiv_1906.06115_2019}. Instead we are interested in Branch I of the stationary solutions ($\alpha \geq 0$) in the presence of a spherically-symmetric trap, that is, $\gamma = 1$. Via Eq.~\eqref{eq:alphaomegadefn} we see that a nonzero $\alpha$ implies that the condensate density profile `rotates' in the laboratory frame despite the velocity field being irrotational and free of vorticity. When the harmonic trapping of a (non)dipolar condensate in the TF limit is subject to a quasi-adiabatic angular acceleration from $\Omega = 0$, the condensate deviates from the expected TF solution at a critical rotation frequency $\Omega = \Omega_{\text{v}}$~\cite{prl_87_19_190402_2001, prl_86_20_4443-4446_2001, pra_73_6_061603r_2006, prl_98_15_150401_2007, pra_80_3_033617_2009} where vortices enter the system and the condensate acquires a nonzero angular momentum. It has also been established in such systems that this vortex state, while energetically favourable to the TF profile at nonzero rotation frequencies, will emerge from the TF solution only in the presence of a \textit{dynamical instability} that forces the system away from the stationary solution. Since the bifurcation diagram shown in Fig.~\ref{fig:bothrangeomegaprofile}(a) is qualitatively similar to that of a (non)dipolar condensate subject to a rotating harmonic trap with a nonzero ellipticity in the $x$-$y$ plane~\cite{prl_86_3_377-390_2001, pra_80_3_033617_2009}, we expect that slowly increasing the rotation frequency of the dipole polarization from zero will result in the triggering of a dynamical instability and a subsequent transition to a state with vortices. We explore this idea via two complementary methods, the first being simulations of the dGPE in the TF regime of a dipolar BEC being subject to an acceleration of the rotation of the dipole polarization about $\hat{z}$, and the second being a linearization of Eqs.~\eqref{eq:continuity} and~\eqref{eq:tfeuler} about the corresponding TF stationary states along Branch I over a range of values of $\edd$.

\begin{figure*}
\includegraphics[width=0.80\linewidth]{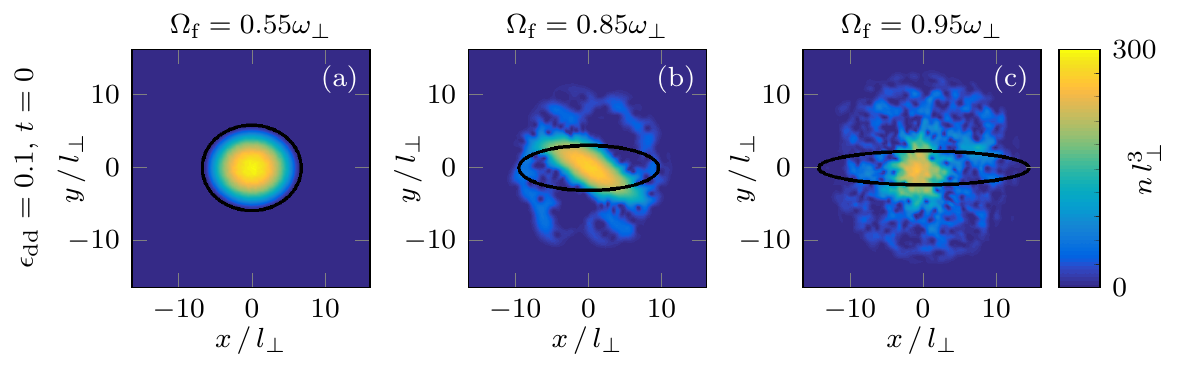}\\
\includegraphics[width=0.80\linewidth]{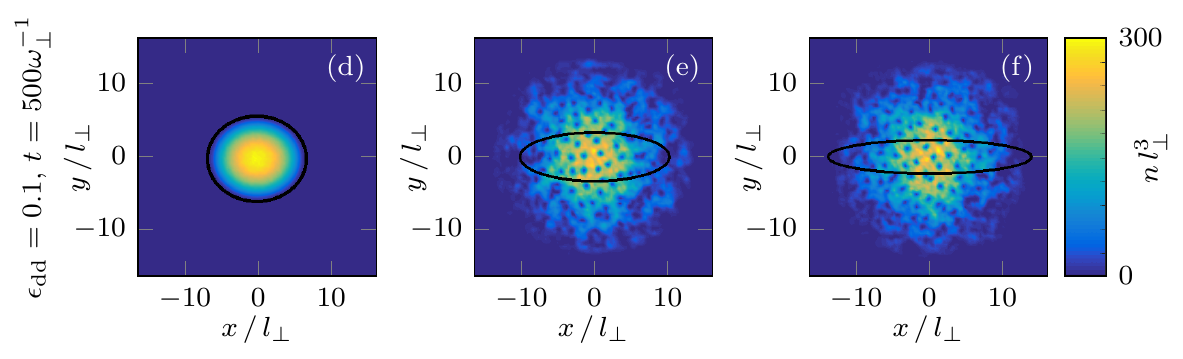}\\
\rule{0.8\textwidth}{.4pt}
\includegraphics[width=0.80\linewidth]{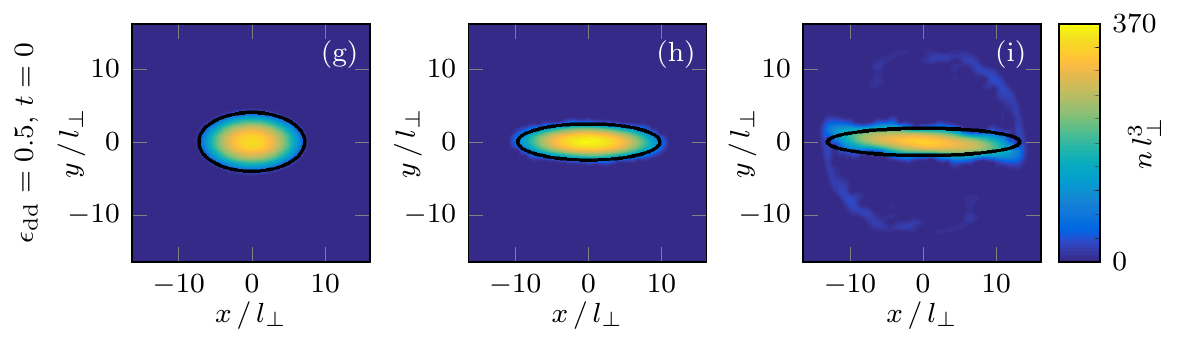}\\
\includegraphics[width=0.80\linewidth]{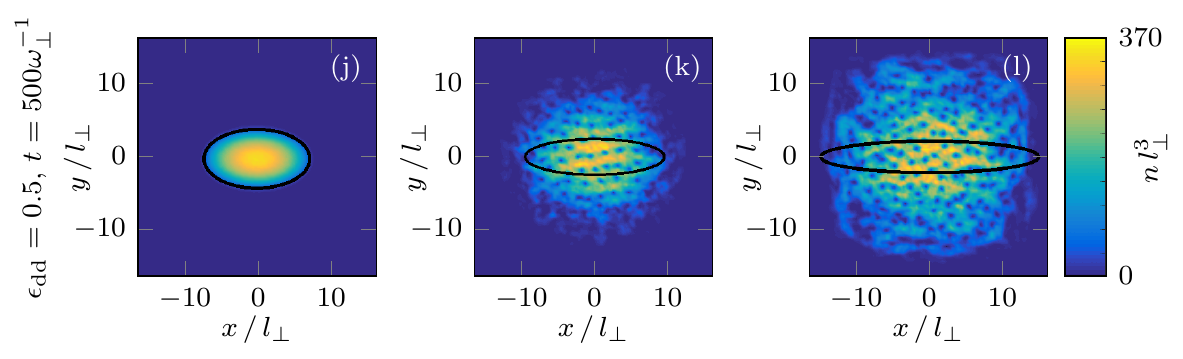}
\caption{Cross-sections at $z = 0$ of density of a dipolar BEC during an quasi-adiabatic ramp-up of $\Omega$ up to $\Omega_\text{f}=(0.55,0.85,0.95)\omega_\perp$. Snapshots taken at $t=0$ in (a)-(c) and (g)-(i) and $t=500\omega_\perp^{-1}$ in (d)-(f) and (j)-(l). In (a)-(f) $\epsilon_\text{dd} = 0.1$ and (g)-(l) $\epsilon_\text{dd}=0.5$. Lengths and density are scaled by $l_{\perp} = \sqrt{\hbar/(m\omega_{\perp})}$. The black ellipses depict the predicted Branch I TF profiles from Eqs.~\eqref{eq:kappax},~\eqref{eq:kappay} and~\eqref{eq:alphakappa}.}
\label{fig:simulation}
\end{figure*}

The numerical integration of Eq.~\eqref{eq:dgpe} is performed with the ADI-TSSP method~\cite{bao_wang_2006}, which is an extension of the split-step Fourier method incorporating rotation. All simulations are undertaken with a  192$^3$ grid with the spatial step $d = 0.15\sqrt{\hbar/(m\omega_{\perp})} \equiv 0.15l_{\perp}$ and temporal step {$\Delta t = 0.004\omega_\perp^{-1}$}. In order to make direct comparisons with the TF analysis, we require a large number of bosons~\cite{pra_78_4_041601r_2008} and thus we fix $N = 10^5$. Since fast Fourier transform algorithms are naturally periodic, we employ a spherical cut-off to the dipolar potential, the second term of Eq.~\eqref{eq:ddpot}, and restrict the range of the DDI to a sphere of radius $R_c$ in real-space in order to reduce the effects of alias copies. Therefore, using the convolution theorem, the interaction potential can be re-expressed as~\cite{ronen_bortolotti_2006}
\begin{align}
V_\text{int}(\textbf{r},t)=g|\psi|^2+\frac{C_\text{dd}}{3}\mathcal{F}^{-1}\left[\tilde{U}_\text{dd}^{R_c}(\textbf{k})\mathcal{F}\left[|\psi|^2\right]\right]\,,
\end{align}
where $\mathcal{F}$ ($\mathcal{F}^{-1}$) denotes the (inverse) Fourier transform and the $\textbf{k}$-space dipolar pseudo-potential is
\begin{align}
\tilde{U}_\text{dd}^{R_c}(\textbf{k}) = \left[1+3\frac{\cos\left(R_ck\right)}{R_c^2k^2}-3\frac{\sin\left(R_ck\right)}{R_c^3k^3}\right]\left(3\cos^2\theta_k-1\right)\,. \label{eq:ddikspacecutoffap}
\end{align}
Here, $\theta_k$ is the angle between $\textbf{k}$ and the direction of the dipoles, and the cut-off radius $R_c$ is chosen to be larger than the system size.

In order to verify the results presented in Fig.~\ref{fig:bothrangeomegaprofile}(a) we take the imaginary time stationary solution for $\Omega=0$ and a fixed $\epsilon_\text{dd}$ and very slowly ramp up $\Omega$ in real time, traversing through the stationary solutions. We employ the following procedure for the real-time ramp rate of $\Omega$,
\begin{align}
\dfrac{\mathrm{d}\Omega}{\mathrm{d}t} = 5\times10^{-4}\,\Theta(\Omega_{\text{f}}-\Omega)\,\omega_{\perp}^2\,,
\end{align}
where $\Theta(\cdot)$ is the Heaviside function and $\Omega_{\text{f}}>0$ is a final constant choice of rotation frequency, at which the acceleration is halted. In this procedure we define $t=0$ to be the time when $\Omega=\Omega_\text{f}$. To assess the stability of these stationary states we employ the above ramp procedure, terminating the angular acceleration at the specified $\Omega_{\text{f}}$, then allowing the system to evolve for $t=500\omega_\perp^{-1}$. Any slow growing instabilities will have had time to seed vortices into the system, allowing us to precisely define a critical $\Omega = \Omega_{\text{v}}$ at the boundary between vorticity and vorticity-free solutions. We also model the the random external symmetry-breaking perturbations that may shift the condensate state away from the stationary state in an experimental scenario by modifying the condensate density at the initial timestep; at each spatial grid point, the condensate density is subjected to a random, local perturbation of up to $5\%$ of the local density. To ensure that the ramp rate is as close to being adiabatic as possible, we have performed several test simulations for half and double the ramp rate, and found a negligible difference for the onset of instability.

Figure~\ref{fig:simulation} depicts the density with fixed $\edd = 0.1$ (panels (a)-(f)) and $\edd=0.5$ (panels (g)-(i)), as cross-sections at $z = 0$ ramped up to rotation frequencies $\Omega_\text{f} = 0.55\omega_\perp$, $0.85\omega_\perp$, and $0.95\omega_\perp$. Odd numbered rows show snapshots taken to $t=0$, i.e. the first moment when $\Omega=\Omega_\text{f}$, and even numbered rows show snapshots after $t=500\omega_\perp^{-1}$ evolution. Overlaid on these density profiles is the predicted TF solution, found through the following procedure: first, the value of $R_z$ is found through a numerical fit of the simulation to a 3D TF profile, then the $(x,y)$ TF radii are generated from Eq.~\eqref{eq:kappax} as $R_x=\kappa_xR_z$ and Eq.~\eqref{eq:kappay} as $R_y=\kappa_yR_z$ from the solutions for $\alpha$ for Branch I. For low $\Omega$, the condensate density is consistent with the TF stationary solution: the density is smooth and approximates the paraboloid profile of Eq.~\eqref{eq:tfdensity}. However, at some critical $\Omega=\Omega_{\text{v}}$ the solution goes through a dynamical instability. This is first visible through a rippling of the density and then the formation of a spiral-like structure that later evolves towards a turbulent vortex state.

\begin{figure}
\includegraphics[width=\columnwidth]{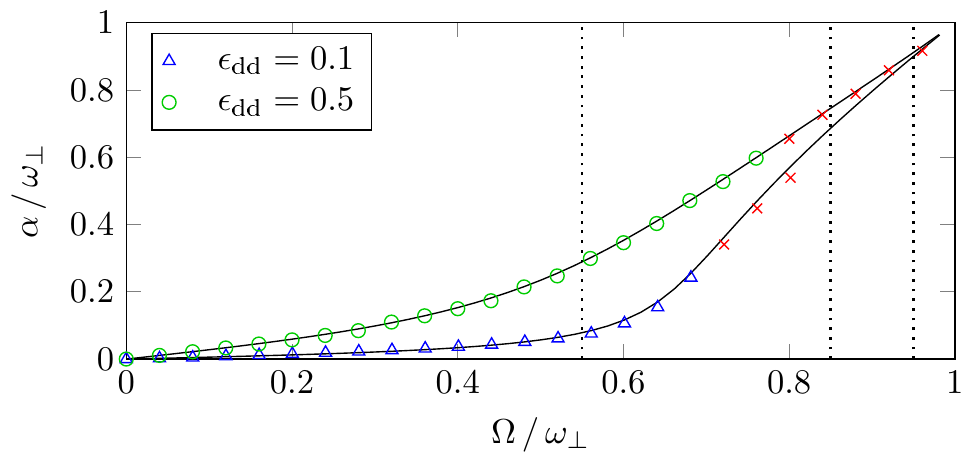}
\caption{Comparison of the TF solutions to the numerical integration of the dGPE. Black lines show the TF predictions for Branch I, underlying the numerically extracted $\alpha$ for $\edd=0.1$ (blue triangles) and $\edd=0.5$ (green circles). Red crosses show solutions that are unstable after a long-time evolution with $\Omega=\Omega_\text{f}$ fixed. Dotted lines are the columns of Fig.~\ref{fig:simulation}.}
\label{fig:comparison}
\end{figure}

With the dGPE simulations in hand, it is also possible to compare the results from these simulations to the predicted density profiles from the TF self-consistency relations in Eqs.~\eqref{eq:kappax},~\eqref{eq:kappay} and~\eqref{eq:alphakappa}. Figure~\ref{fig:comparison} compares the numerically obtained solutions for $\alpha$ (markers) with those from the TF analysis (solid lines), for the same parameters as those shown in Fig.~\ref{fig:simulation}. All values of $\alpha$ are measured at the end of the ramp procedure, at $t=0$, however the red crosses show simulations where solutions are unstable at long times (c.f.~panels (h) and (k)). Quite remarkably, the larger $\edd$ solution is dynamically stable for faster $\Omega$. For $\edd=0.1$ and $\Omega>0.8\omega_\perp$ vortices enter the condensate and there is no sensible measure of $\alpha$. The boundary between the red crosses and the blue triangles or green circles is indicative of the value of $\Omega_{\text{v}}$ for each $\edd$.

\begin{figure*}
\includegraphics[width=0.80\linewidth]{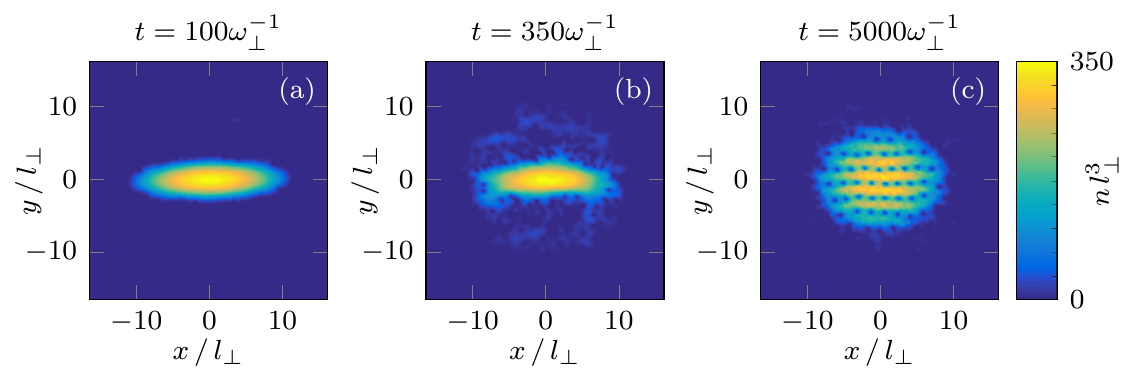}
\caption{Long-time evolution of a dipolar BEC after a ramp up to $\Omega=\Omega_\text{f}$, with $\Omega_{\text{f}}=0.850\omega_\perp$ and $\edd=0.5$, into a triangular vortex lattice.}
\label{fig:vlatt}
\end{figure*}

We also find that, given sufficient time to evolve, the vortices we observe in the dGPE simulation form a vortex lattice. This is shown in Fig.~\ref{fig:vlatt} for $\edd=0.5$ and $\Omega_{\text{f}}=0.850\omega_\perp$, taking the initial condition from Fig.~\ref{fig:simulation}(h), after a total temporal evolution of $t=5000\omega_\perp^{-1}$. After $t=100\omega_\perp^{-1}$ the condensate density shows small fluctuations on its surface and at $t=350\omega_\perp^{-1}$ large quantities of vortices have entered the condensate, but are still confined to the peripheries of the density. Finally, by $t=5000\omega_\perp^{-1}$ the vortices have entered the condensate and have relaxed into a triangular Abrikosov vortex lattice. Several theoretical studies have indicated that this may be the ground state for a system of vortices in (non-)dipolar BECs at finite rotation frequencies~\cite{prl_95_20_200402_2005, prl_95_20_200403_2005, pra_75_2_023623_2007, rmp_81_2_647-691_2009, pra_98_2_023610_2018, jphysb_49_15_155301_2016, jphyscondesmatter_29_10_103004_2017}. In nondipolar BECs subject to a ramping up of the trapping rotation frequency, a vortex lattice is seen to form following a period of evolution of the nondipolar GPE at constant rotation frequency after the vortex instability has been triggered~\cite{prl_92_2_020403_2004, pra_73_6_061603r_2006}. In the context of dipolar BECs, the results of $3$D study with no contact interactions, i.e. $a_s = 0, \epsilon_{\text{dd}} \rightarrow \infty$, also predict a square condensate density like that of Fig.~\ref{fig:simulation}(l)~\cite{jphysb_49_15_155301_2016}.

However, a comprehensive numerical dGPE study of the ground state configurations of vortex lattices in quasi-$2$D trapping geometries, i.e. $\gamma \gg 1$, finds that when the dipole orientation is in the $x$-$y$ plane, a triangular lattice configuration is favoured for $\edd \lesssim 0.8$ and a striped condensate phase with a square lattice configuration is favourable for $\edd \gtrsim 0.8$ in the $\Omega\rightarrow\omega_{\perp}$ limit~\cite{pra_98_2_023610_2018}. Given that a triangular vortex lattice is predicted for the regimes that we consider in our work, albeit in a quasi-$2$D geometry, it is possible that a striped condensate phase with a square vortex lattice geometry may be seen if we allow $\Omega_{\text{f}}$ to approach $\omega_{\perp}$ and if $\edd \rightarrow 1$. However, the dimensionality of the system would be expected to play a significant role in the nature of the vortex lattice configuration, and beyond-mean-field effects play a significant role in the behaviour of dipolar BECs with large $\edd$~\cite{pra_86_6_063609_2012, pra_94_3_033619_2016}. As such, a direct comparison of our methodology with these results cannot be made. A thorough consideration of beyond-mean-field effects as well as the eventual vortex lattice configurations across different regimes of $\edd$ and $\Omega_{\text{f}}$ is beyond the scope of this work, but still warrants further study.

\section{\label{sec:level4}Linearization of the Thomas-Fermi Stationary Solutions}
The solutions of the dGPE equation clearly depict the emergence of vortices in the vorticity-free TF stationary states after a certain critical rotation frequency, $\Omega = \Omega_{\text{v}}(\edd)$. As each real-time evolution of the dGPE begins with a random symmetry-breaking perturbation of the density, a natural conclusion would be that this perturbation induces a dynamical instability of the condensate that allows for a dynamical route from the TF stationary states to lower energy states such as vortex states at high rotation frequencies~\cite{prl_87_19_190402_2001, pra_80_3_033617_2009}. In order to understand when the stationary states become dynamically unstable, we linearize the fully time-dependent hydrodynamic equations, Eqs.~\eqref{eq:continuity} and~\eqref{eq:tfeuler}, about the TF stationary states as predicted by Eq.~\eqref{eq:continuitystat} and~\eqref{eq:tfeulerstat}. This yields a formulation of the fully time-dependent state in terms of collective modes about the stationary states, one or more of which may be dynamically unstable against external perturbations. Initially, we express the time-dependent solutions to Eq.~\eqref{eq:continuity} and~\eqref{eq:tfeuler} as
\begin{align}
n(\mathbf{r}, t) &= n_{\text{TF}}(\mathbf{r}) + \delta n(\mathbf{r}, t), \label{eq:densitypert} \\
S(\mathbf{r}, t) &= S_{\text{TF}}(\mathbf{r}, t) + \delta S(\mathbf{r}, t). \label{eq:phasepert}
\end{align}
Here, $n_{\text{TF}}(\mathbf{r})$ and $S_{\text{TF}}(\mathbf{r}, t)$ are given by Eqs.~\eqref{eq:tfdensity} and~\eqref{eq:tfphase}, thus satisfying Eqs.~\eqref{eq:continuitystat} and~\eqref{eq:tfeulerstat}, while $\delta n$ and $\delta S$ are considered to be `small' perturbations about the stationary states. The linearization of the time-dependent problem is subsequently carried out by substituting Eqs.~\eqref{eq:densitypert} and~\eqref{eq:phasepert} into Eqs.~\eqref{eq:continuity} and~\eqref{eq:tfeuler}, utilising Eqs.~\eqref{eq:continuitystat} and~\eqref{eq:tfeulerstat} to simplify the result, and neglecting contributions from terms that are of higher order than linear in the fluctuations $\delta n$ and $\delta S$. The resulting system of coupled first-order ordinary differential equations is given by~\cite{prl_87_19_190402_2001, pra_73_6_061603r_2006, prl_98_15_150401_2007, pra_80_3_033617_2009}:
\begin{gather}
\frac{\partial}{\partial t}
\begin{pmatrix}
\delta S \\
\delta n
\end{pmatrix}
=
\mathbb{M}
\begin{pmatrix}
\delta S \\
\delta n
\end{pmatrix}, \label{eq:perteqns} \\
\mathbb{M} = -\begin{pmatrix}
\mathbf{v}_c\cdot\nabla & \frac{g}{\hbar}\left(1-\edd\widehat{K}\right) \\
\frac{\hbar}{m}\nabla\cdot\left(n_{\text{TF}}\nabla\right) & \mathbf{v}_c\cdot\nabla
\end{pmatrix}, \label{eq:pertmatrix} \\
\mathbf{v}_c = \frac{\hbar}{m}\nabla S_{\text{TF}} - \mathbf{\Omega}\times\mathbf{r}, \label{eq:labvel} \\
\widehat{K}\left[\delta n\right] = \delta n + 3\frac{\partial^2}{\partial x^2}\int_{\Gamma_{\text{TF}}}\mathrm{d}^3r'\,\frac{\delta n\left(\mathbf{r}',t\right)}{4\pi\left|\mathbf{r}-\mathbf{r}'\right|}. \label{eq:koperator}
\end{gather}
Here $\Gamma_{\text{TF}}$ is defined as $\Gamma_{\text{TF}} = \lbrace\mathbf{r}\in\mathbb{R}^3:n_{\text{TF}}(\mathbf{r}) > 0\rbrace$. The linearized fluctuations are given by a sum of collective modes, indexed by $\nu$, such that
\begin{gather}
\begin{pmatrix}
\delta S(\mathbf{r}, t) \\
\delta n(\mathbf{r}, t)
\end{pmatrix}
=
\sum_{\nu}
\begin{pmatrix}
\delta S_{\nu}(\mathbf{r}) \\
\delta n_{\nu}(\mathbf{r})
\end{pmatrix}
\exp(\lambda_{\nu} t), \label{eq:collmodes} \\
\mathbb{M}
\begin{pmatrix}
\delta S_{\nu} \\
\delta n_{\nu}
\end{pmatrix}
=
\lambda_{\nu}
\begin{pmatrix}
\delta S_{\nu} \\
\delta n_{\nu}
\end{pmatrix}. \label{eq:eigproblem}
\end{gather}

To diagonalize Eq.~\eqref{eq:eigproblem}, we use a monomial basis for $\delta n_{\nu}$ and $\delta S_{\nu}$ of the form $\lbrace x^iy^jz^k\rbrace$~\cite{prl_87_19_190402_2001}. By inspection of $\mathbb{M}$, a given mode will feature the same value of $l =\max{[i + j + k]}$ for both $\delta n$ and $\delta S$, a number which we refer to as the order of the mode. While it is relatively simple to calculate how a given monomial is transformed by the nondipolar components of $\mathbb{M}$, the evaluation of the dipolar contribution is quite involved, but by using methods originally developed for the study of Newtonian potentials inside classical self-gravitating ellipsoidal fluids, it is possible to calculate $\widehat{K}\left[\delta n\right]$~\cite{apj_166_441-445_1971, pra_82_3_033612_2010, pra_82_5_053620_2010}. For a polynomial $x^iy^jz^k$, we rewrite the exponents as
\begin{equation}
i = 2\lambda + \delta_{\lambda}\,,\, j = 2\mu + \delta_{\mu}\,,\, k = 2\nu + \delta_{\nu}, \label{eq:ijkdecomp}
\end{equation}
where $\lbrace\delta_{\lambda}, \delta_{\mu}, \delta_{\nu}\rbrace \in \lbrace 0, 1\rbrace$ and $\lbrace\lambda, \mu, \nu\rbrace$ are integers. We also introduce the definition $\sigma = \lambda + \mu + \nu + 1$. The integral in $\mathbb{M}$ is then given by
\begin{widetext}
\begin{equation}
\int_{\Gamma_{\text{TF}}}\frac{x'^iy'^jz'^k\,\mathrm{d}^3r'}{4\pi\left|\mathbf{r}-\mathbf{r}'\right|} = \frac{R_x^iR_y^jR_z^ki!j!k!}{2^{2\sigma-1}}\sum_{p=0}^{\sigma}\sum_{q=0}^{\sigma-p}\sum_{r=0}^{\sigma-p-q}\frac{(-2)^{p+q+r}x^{2r+\delta_{\lambda}}y^{2q+\delta_{\mu}}z^{2p+\delta_{\nu}}\Lambda_{pqr}^{(i,j,k)}}{(2p)!(2q)!(2r)!(\sigma-p-q-r)!(2r\delta_{\lambda}+1)(2q\delta_{\mu}+1)(2p\delta_{\nu}+1)}, \label{eq:dipolekernelaction}
\end{equation}
\end{widetext}
where the expression is simplified by the definitions
\begin{widetext}
\begin{gather}
\Lambda_{pqr}^{(i,j,k)} = \sum_{l=0}^{\lambda}\sum_{m=0}^{\mu}\sum_{n=0}^{\nu}\frac{(-2)^{p+q+r}R_x^{2l+\delta_{\lambda}}R_y^{2m+\delta_{\mu}}R_z^{2n+\delta_{\nu}}M_{l+p+\delta_{\lambda},m+q+\delta_{\mu},n+r+\delta_{\nu}}}{(2p)!(2q)!(2r)!(\lambda-l)!(\mu-m)!(\nu-n)!(2l\delta_{\lambda}+1)(2m\delta_{\mu}+1)(2n\delta_{\nu}+1)}, \label{eq:lambdakernelcoefficient} \\
M_{lmn} = (2l-1)!!(2m-1)!!(2n-1)!!\frac{\kappa_x\kappa_y\beta_{lmn}}{2R_z^{2(l+m+n-1)}}, \label{eq:depolcoeff}
\end{gather}
\end{widetext}
and $\beta_{ijk}$ represents the integrals defined in Eq.~\eqref{eq:beta}. The dipolar contribution to $\mathbb{M}$ is obtained by taking the second partial derivative of the expression on the RHS of Eq.~\eqref{eq:dipolekernelaction} with respect to $x$. We also note that, formally, Eq.~\eqref{eq:pseudopotbeta} may be derived via use of Eqs.~\eqref{eq:dipolekernelaction} --~\eqref{eq:depolcoeff}.

After diagonalizing Eq.~\eqref{eq:eigproblem}, the TF stationary state is said to be \textit{dynamically unstable} if any mode $\nu$ is characterised by $\text{Re}[{\lambda_{\nu}}]> 0$, since the magnitude of the mode grows exponentially with time. Within this framework, we may be certain that a stationary state is stable only if there are no eigenvalues $\lambda_{\nu}$ possessing a positive real component. As there are an infinite-number of possible collective modes, it is necessary to fix a truncation parameter, $N_{\text{max}}$, such that $l = \max{[i + j + k]} \leq N_{\text{max}}$. For a choice of $N_{\text{max}}$, it is possible that there are modes of a higher order at a given point in parameter space that are unstable and, as such, the absence of unstable modes for a given $N_{\text{max}}$ is not a guarantee of dynamical stability. However, if a sensible choice of $N_{\text{max}}$ yields at least one unstable mode, it is sufficient to claim that the corresponding stationary state is dynamically unstable. Due to computational constraints we fix $N_{\text{max}} = 10$, a choice which we consider to be sufficient for a qualitative analysis. We also note that the $l = 0$ and $l = 1$ modes are always dynamically stable. Firstly, by inspection of $\mathbb{M}$, it is clear that truncating the basis to $N_{\text{max}} = 0$ yields two null eigenvalues. As for the $l = 1$ \textit{dipole} modes, which represent centre-of-mass rigid-body oscillations of the condensate, these are exactly decoupled from the two-body interactions of the condensate as per Kohn's theorem: the eigenvalues of the six dipole modes are purely imaginary and are given by $\text{Im}(\lambda_{l = 1})\in\lbrace\omega_{\perp}\pm\Omega, \gamma\omega_{\perp}\rbrace$, along with their complex conjugates~\cite{prl_86_3_377-390_2001}.

\begin{figure}
\begin{center}
  \begin{tabular}{c c}
    \begin{minipage}[c][0.44\columnwidth][c]{0.498\columnwidth}
    \centering
      \includegraphics[width=\columnwidth]{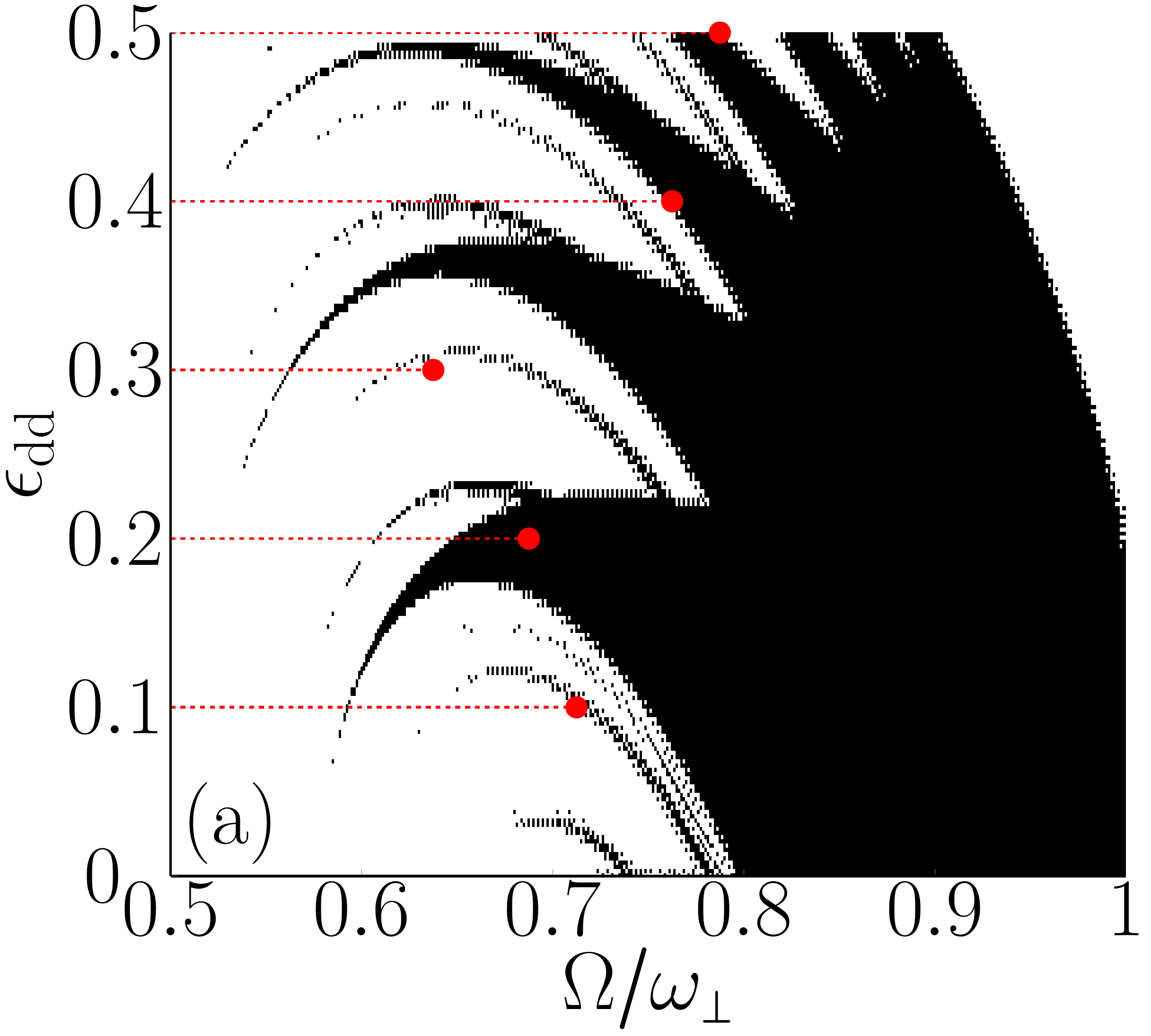}
    \end{minipage}
    &
    \begin{minipage}[c][0.44\columnwidth][c]{0.498\columnwidth}
    \centering
    \includegraphics[width=\columnwidth]{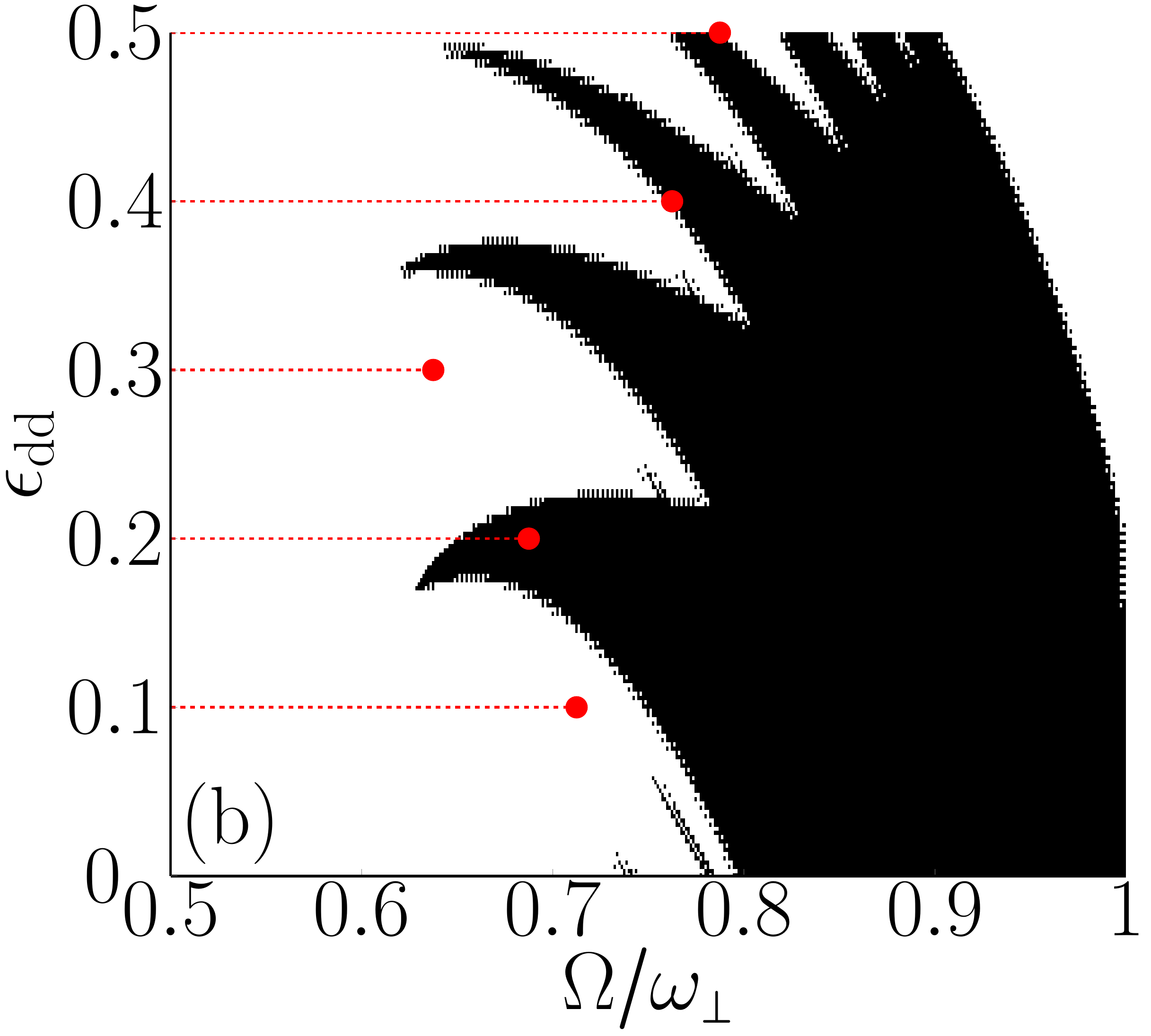}
    \end{minipage}
  \end{tabular}
\end{center}
\vspace*{-0.5cm}
\caption{Dynamical instability of the TF stationary states: With the truncation parameter $N_{\text{max}} = 10$, the shaded region indicates where there exists at least one linearized eigenvalue with a positive real component greater than $10^{-4}$ (a) or $10^{-2}$ (b). Overlain in red are the trajectories of the dGPE simulations for $\edd\in\lbrace 0.1, 0.2, 0.3, 0.4, 0.5\rbrace$ with filled circles at $\Omega_{\text{v}}$.}
\label{fig:heavisidemaps}
\end{figure}

Let us consider the domain $\Omega/\omega_{\perp}\in[0.5, 1]$ and $\edd \in [0, 0.5]$, which roughly corresponds to the domain in which the vortex instability was observed in the dGPE simulations. In Fig.~\ref{fig:heavisidemaps}, points in parameter space are shaded if we find at least one eigenvalue of $\mathbb{M}$ with a positive real component greater than $10^{-4}$ (a) and $10^{-2}$ (b). Figure~\ref{fig:heavisidemaps} demonstrates that, at higher values of $\Omega < \omega_{\perp}$, the Branch I TF stationary solutions become dynamically unstable. The resulting region of dynamical instability is found to be continuous as $\Omega\rightarrow\omega_{\perp}$ but separating into distinct arcs for lower $\Omega$. Each of these arcs corresponds to distinct polynomial orders $l \geq 3$ and is connected to the bulk of the dynamically unstable region at successively higher values of both $\Omega$ and $\edd$. There is also an instability of an $l = 2$ mode, corresponding to a narrow arc that is disconnected from the connected region of instability and emerges at $\edd = 0$ at $\Omega/\omega_{\perp} \approx 0.74$. In Fig.~\ref{fig:heavisidemaps} (a), we also observe additional narrow arcs of instability corresponding to slowly growing high-order unstable modes. These are not present in Fig.~\ref{fig:heavisidemaps}(b) due to the higher eigenvalue cutoff. We note that for larger $\edd$, Figs.~\ref{fig:heavisidemaps}(a) and (b) imply dynamical stability near $\Omega = \omega_{\perp}$, but this is merely due to the choice of $N_{\text{max}}$ being insufficiently large in order to accurately probe the dynamical stability of the solutions in this regime. However, Fig.~\ref{fig:heavisidemaps} is qualitatively similar to the corresponding stability diagrams for both nondipolar BECs and $z$-polarized dipolar BECs in asymmetrical rotating harmonic traps, where the transverse trapping ellipticity plays an analogous role to $\edd$ in our system~\cite{prl_87_19_190402_2001, prl_98_15_150401_2007, pra_80_3_033617_2009}. In these analogous systems, for a given ellipticity, the stationary solutions are always unstable above a critical rotation frequency and so we expect that the solutions are dynamically unstable when $\Omega\rightarrow\omega_{\perp}$, a property that would be clear if a larger value of $N_{\text{max}}$ had been chosen.

We now proceed to compare the predictions of the dGPE simulations and the linearization of the TF hydrodynamic equations regarding the dynamical instability of the condensate. Overlain on Fig.~\ref{fig:heavisidemaps} are red dashed lines at constant $\edd$, each representing the trajectories of dGPE simulations utilising ramp procedure $(2)$ as described in Section~\ref{sec:level3}. Each dashed line terminates in a red star at $\Omega = \Omega_{\text{v}}(\edd)$, the rotation frequency which we estimate to be the lowest rotation frequency at which the dynamical instability develops from an average over $10$ simulations. For low $\edd$ we find that $\Omega_{\text{v}}$ corresponds roughly to the narrow, high-order modes seen in Fig.~\ref{fig:heavisidemaps}(a), while the bulk of the unstable region of parameter space that survives in Fig.~\ref{fig:heavisidemaps}(b) seems to be responsible for the dynamical instability when $\edd$ equals $0.4$ or $0.5$. This is a somewhat unusual departure from previous analyses of similar systems, such as the rotating asymmetrically trapped condensate, as the instabilities in these systems seem to be driven exclusively by the bulk of the instability region rather than the narrow instability arcs in the corresponding eigenvalue plots~\cite{prl_87_19_190402_2001, prl_98_15_150401_2007, pra_80_3_033617_2009}. This tendency for the condensate to be more unstable than expected warrants further study.

\section{\label{sec:level5}Conclusion}
We have identified a new mechanism for the dynamical formation of vortex lattices in dipolar BECs, a phenomenon which has previously been considered in the context of time-dependent rotating traps. Using the dGPE and the dipolar superfluid hydrodynamic equations in the Thomas-Fermi limit, we have investigated the effects of a slow increase of the rotation frequency of the dipole moments about the cylindrical axis of the applied harmonic trap. Simulations of the dGPE demonstrate that the stationary solutions in the TF regime suffer a dynamical instability that causes the dipolar condensate to spontaneously reduce its energy and acquire a nonzero angular momentum via the seeding of vortices. Further time evolution of the dGPE reveals that the the vortices self-order into a triangular Abrikosov vortex lattice, a ground state that has been predicted for a dipolar BEC in numerous prior studies. The domain of parameter space that supports such a dynamical instability is further explored via linearization of the dipolar superfluid hydrodynamic equations about the TF stationary states at a given rotation frequency and dipolar coupling strength. By characterizing the linear stability of the collective modes obtained via this procedure, we find that a dynamical route to the vortex lattice ground state is available for any value of the dipolar interaction strength as the rotation frequency approaches the transverse trapping frequency. This result paves the way to studying vortex lattices in dipolar BECs in experiments without any modification to the externally applied trap, but instead via manipulation of the orientation of the dipole moments themselves. In particular, such systems may harbor novel phases such as a square vortex lattice phase~\cite{prl_95_20_200402_2005, prl_95_20_200403_2005, jphyscondesmatter_29_10_103004_2017}. The technology to realise this scheme experimentally already exists at the time of writing, as the polarization of a dipolar BEC has been rotated at considerably higher rotation frequencies in an experiment investigating the effective rotational tuning of the dipolar interaction in the rapid rotation limit~\cite{prl_120_23_230401_2018}. While we predict the formation of a triangular vortex lattice in the parameter regimes we have explored, it may be possible to obtain square lattice configurations for $\edd \sim 1$ and this, as well as the effects of modifying the trapping aspect ratio, $\gamma$, warrants further study. Finally, we note that the mechanism detailed here may be generalized by considering the rotation of dipole moments tilted away from the $x$-$y$ axis, which might alter the nature of the vortex instability and the properties of the resulting vortex lattice.

\begin{acknowledgments}
S.B.P. is supported by an Australian Government Research Training Program Scholarship and by the University of Melbourne. A.M.M. would like to thank the Institute of Advanced Study (Durham University, U.K.) for hosting him during the initial stages of developing this collaborative research project and the Australian Research Council (Grant No. LE180100142) for support. T.B. and N.G.P. thank the Engineering and Physical Sciences Research Council of the UK (Grant No. EP/M005127/1) for support.
\end{acknowledgments}

\bibliography{main.bbl}
\end{document}